\documentstyle[aps]{revtex}

\begin{document}

\begin{center}
{\Large Dimerization of Ferrimagnets on Chains and Square Lattices}

Aiman Al-Omari\footnote{%
aiman@physics.sdnpk.undp.org} and A. H. Nayyar\footnote{%
nayyar@physics.sdnpk.undp.org}

Department of Physics, \\[0pt]Quaid-i-Azam University, \\[0pt]Islamabad
45320, Pakistan.

\vspace{.7cm}( 1998)\vspace{1.2cm}

{\bf Abstract}\vspace{0.5cm}
\end{center}

A linear spin wave analysis of dimerization of alternating Heisenberg system
with spins $s_1$ and $s_2$ on linear chain as well as square lattice is
presented. Among the several possible dimerized configurations considered in
two dimensions the plaquette configuration is found to be energetically the
most favored one. Inclusion of a variable nearest neighbor exchange coupling 
$J(a)=\frac Ja$ leads to a uniform power law behavior: that is to say, the
same $\delta $-dependence is found (i) in chains as well as in square
lattices; (ii) in systems consisting of different pairs of spins $s_1$ and $%
s_2$; (iii) for the magnetic energy gain, the energy gap, the energy of the
gapped magnetic excitation mode as well as for the sublattice magnetization;
(iv) for all the configurations of the square lattice; and (v) in the entire
range of $\delta :$ $(0\leq \delta <1)$. The variable exchange coupling also
allows \ the energy of the gapped excitation spectrum to be $\delta $%
-dependent even in the linear spin wave theory.\vspace{0.8cm}

\vspace{0.5cm}\newpage

\section{Introduction:}

Extensive interest is being currently shown in alternating spin systems
consisting of two sublattices with unequal spin magnitudes $s_{1\text{ }}$%
and $s_2$ with a net non-zero spin per unit cell, as shown in Fig.1. Such
systems are realized in bi-metallic chains with the general formula of ACu(%
{\em pba}OH)(H$_2$O)$_3$.2H$_2$O where {\em pba} OH is
2-hydroxy-1,3-propylenebis (oxamato) and A = Mn, Fe, C, Ni\cite{petra}.
These ferrimagnetic chains are also referred to as alternating or mixed spin
chains and are regarded as Heisenberg systems\cite
{drillon,kolezhuk,brehmer,swapan,pati}.\newline

Alternating spin systems have been studied extensively by various
techniques: by spin wave theory (SWT)\cite
{brehmer,swapan,pati,blote,yamamoto,yamamoto2,yamamoto3}, spin wave
expansion (SWE)\cite{ivanov,ivanov2}, Monte Carlo (MC) method \cite
{brehmer,yamamoto,yamamoto2,yamamoto3}, density matrix renormalization group
(DMRG) technique\cite{swapan,yamamoto3,ivanov2}, the method of matrix
product (MP) states\cite{kolezhuk} and by exact diagonalization (ED) method%
\cite{yamamoto2,ivanov2}. \\

For an alternating-spin chain the zero temperature ground state energy and
sublattice magnetization were evaluated using SWT\cite{pati,swapan,yamamoto3}%
, SWE\cite{ivanov}, DMRG \cite{swapan,pati,yamamoto3} and QMC\cite{kolezhuk}%
. The results are summarized in Table 1. The linear spin wave theory gives
higher values for the ground state energy and lower values for the
sub-lattice magnetization compared to the more exact methods. Recently
Ivanov {\em et al.}\cite{ivanov,ivanov2} used a second-order spin wave
expansion to calculate ground state energy as well as sublattice
magnetization. Their results differ by 0.03\% for the ground state energy
and 0.2\% for the sublattice magnetization from the DMRG results, as shown
in Table 1.\\

The thermal behavior was also investigated for ferrimagnetic chains\cite
{swapan,pati,yamamoto,yamamoto2,yamamoto3}. Besides verifying the existence
of two (gapped and gapless) excitation modes, the specific heat and magnetic
susceptibility of ferrimagnetic chains were also shown to depend upon
temperature as T$^{1/2}$ and T$^{-2}$ respectively at low temperatures\cite
{yamamoto,yamamoto3}. It was also shown that this model behaved as a
ferromagnet at low temperature, but as a gapped antiferromagnet at moderate
temperatures.

Modified spin wave theory, which includes Takahashi constraint, was also
shown to give results in surprisingly good agreement with those from quantum
Monte Carlo method in the thermodynamic limit of this system\cite
{yamamoto,yamamoto3}.\newline

Dimerization of chains with spins $s_1$ and $s_2$ ( $s_1$ $>$ $s_2$) on
alternating sites was recently \cite{swapan,pati} studied the using the
Hamiltonian 
\begin{equation}
H=J\sum_n\left[ \left( 1+\delta \right) S_{1,n}\cdot S_{2,n}+\left( 1-\delta
\right) S_{2,n}\cdot S_{1,n+1}\right] ,  \label{alt1d}
\end{equation}
where the total number of sites (or bonds) is 2N and the sum is over the
total number of unit cells N. $\delta $ is the dimerization parameter and is
taken to vary between $0$ and $1$. Linear spin wave theory and DMRG were used%
\cite{swapan,pati} to investigate the ground and low-lying excited states
for both uniform and dimerized chains. In both the methods the ground state
was found to be ferrimagnetic. One point of focus for us in the study of
chains is that the LSW theory with the Hamiltonian in Eq.(\ref{alt1d})
showed that the energy gap at $k=0$ in the gapped mode did not depend on the
dimer parameter $\delta ,$ while the DMRG predicted an almost linear
dependence\cite{swapan,pati}.The DMRG results on chains also show that the
transition to a spin-Peierls state is conditional in that the ground state
energy depends upon the dimerization parameter as $\delta ^\nu $ with $\nu $%
=2 $\pm $.01.\newline

This has motivated us to investigate a dimerized alternating spin Heisenberg
model by using a linear spin wave theory using an ansatz of a variable
nearest neighbor exchange coupling that was recently used to study
dimerization in a uniform (single spin) Heisenberg system\cite{aiman}. We
would also like to extend our work to square lattices.\newline

A need for describing nearest neighbor exchange interaction as dimerization
sets in two dimensional lattices was recently discussed\cite{aiman}. Among
the various lattice deformation modes which allow for dimerization, some
require that the effect of the ensuing elongation or contraction of nearest
neighbor distances be reflected in the nearest neighbor spin-spin exchange
couplings. Since an exchange integral for a nearest neighbor distance $a$ is
roughly\cite{comment} 
\begin{equation}
J(a)=\frac Ja,  \label{fullj}
\end{equation}
we assume that when the nearest neighbor distances change from $a$ to $%
a(1\pm \delta )$, the exchange couplings change from $J$ to $\frac J{1\pm
\delta }$. Thus, to linear order in $\delta $, the interaction $\frac J{1\pm
\delta }$ has the familiar form $J(1\pm \delta )$. The form in Eq.(\ref
{fullj}) allows for incorporating changes in the nn exchange couplings in
various situations of lattice deformations. It was shown\cite{aiman} that
the logarithmic $\delta $-dependence of various quantities like the gain in
the magnetic ground state energy, etc., $\varepsilon _g\symbol{126}\frac{%
\delta ^\nu }{\left| \ln \delta \right| },$ both in one and two dimensional
lattices, can also be a result of using this variable nearest neighbor
spin-spin exchange coupling. The form in Eq.(\ref{fullj}) gives a
logarithmic dependence over not just the $\delta \rightarrow 0$ regime, but
over the entire range of $\delta $ from 0 to 1$.$ In what follows, we shall
use for exchange interaction the form in Eq.(\ref{fullj}).\newline

In this paper we will study alternating spin systems formed with different
pairs of spin values; $\frac 12,1,$ and $\frac 32$ using a zero temperature
linear spin wave theory. We have considered three alternating spin systems
from these spin values: $(1,\frac 12)$ (denoting $s_1=1$ and $s_2=\frac 12)$%
; $(\frac 32,\frac 12)$ and $(\frac 32,1)$. We would like to see the effect
of including the variable nearest neighbor exchange coupling on the $\delta $%
-dependence of the physical quantities such as the gain in magnetic energy,
the sub-lattice magnetization and energies of the excitation modes in both
one and two space-dimensions and for different spin systems. We would also
like to see if it gives a $\delta $-dependence of the gapped excitation
energy mode. In section two we will study these three alternating spin
systems for a Heisenberg linear chain using the variable nearest neighbor
exchange coupling. The energy and magnetization of such systems will be
computed using LSW theory. Critical exponents of the dimer alternating
chains will also be calculated. We shall then study alternating spin systems
on a square lattice for several proposed dimer configurations in section III.%
\newline
\label{111}

The question of frustration in a ferrimagnetic chain or a ladder due to an
antiferromagnetic second neighbor interaction has also received some
attention recently. It was shown that a strong frustration leads to
disappearance of the long range ferrimagnetic order through a discontinuous
transition to a singlet state\cite{ivanov2,kolezhuk2,koga}. It was also
shown that the spin wave theory can predict realistic results for a
frustrated system at least for the case of weak frustration\cite{ivanov2}.
We shall attempt to see the effect of the variable nearest neighbor exchange
interaction on weakly frustrated chains and square lattices in a future
publication.

\section{ One Dimensional Alternating System}

The alternating dimer Hamiltonian on a chain with two spins $s_1$ and $s_2$
can be rewritten using the variable nearest neighbor exchange coupling
defined in Eq.(\ref{fullj}) as 
\begin{equation}
H=\sum_i\left[ \frac J{1+\delta }S_{1,2i}\cdot S_{2,2i+1}+\frac J{1-\delta }%
S_{2,2i+1}\cdot S_{1,2i+2}\right]  \label{alt}
\end{equation}

A linear spin-wave analysis is usually performed with the help of
Holstein-Primakoff (HP)\ transformations to bosonic spin-deviation
operators. For the two sublattices the HP transformations are: for spin-$s_1$
\begin{mathletters}
\label{hp1}
\begin{eqnarray}
S_{1,n}^{+} &=&(2s_1-a_n^{\dagger }a_n)^{1/2}a_n \\
S_{1,n}^{-} &=&a_n^{\dagger }(2s_1-a_n^{\dagger }a_n)^{1/2} \\
S_{1,n}^z &=&s_1-a_n^{\dagger }a_n
\end{eqnarray}
and for the second sublattice with spin-$s_2$ 
\end{mathletters}
\begin{mathletters}
\label{hp2}
\begin{eqnarray}
S_{2,n}^{+} &=&b_n^{\dagger }(2s_2-b_n^{\dagger }b_n)^{1/2} \\
S_{2,n}^{-} &=&(2s_2-b_n^{\dagger }b_n)^{1/2}b_n \\
S_{2,n}^z &=&b_n^{\dagger }b_n-s_2
\end{eqnarray}
where $s_i$ is the magnitude of the spin on sublattice $i$. A linearized
Hamiltonian is obtained by substituting HP transformations into Eq.(\ref{alt}%
) and keeping terms up to the quadratic order in the spin-deviation
operators $a$ and $b.$ The linearized Hamiltonian in Fourier transformed
variables is

\end{mathletters}
\begin{equation}
H=\sum_k\left[ A_1a_k^{\dagger }a_k+A_2b_k^{\dagger }b_k+B(k)\left(
a_k^{\dagger }b_k^{\dagger }+b_ka_k\right) +C\right]  \label{linh}
\end{equation}
with 
\begin{mathletters}
\label{mom}
\begin{eqnarray}
A_1 &=&J_ps_2 \\
A_2 &=&J_ps_1 \\
B(k) &=&\Lambda _k\sqrt{s_1s_2} \\
C &=&-J_ps_1s_2.
\end{eqnarray}

Here 
\end{mathletters}
\begin{eqnarray}
\Lambda _{k} &=&\sqrt{\left( J_{p}\cos (k)\right) ^{2}+\left( J_{m}\sin
(k)\right) ^{2}} \\
J_{p} &=&\frac{J}{1-\delta ^{2}}  \label{jp}
\end{eqnarray}

and 
\begin{equation}
J_{m}=J_{p}\delta
\end{equation}
.

The linearized Hamiltonian in Eq.(\ref{linh}) can be diagonalized using
Bogoliubov transformations 
\begin{mathletters}
\label{mom}
\begin{eqnarray}
&&a_k=u_k\alpha _k+v_k\beta _k^{\dagger } \\
&&b_k=u_k\beta _k+v_k\alpha _k^{\dagger }
\end{eqnarray}
to

\end{mathletters}
\begin{equation}
\tilde{H}=\sum_{k}\left[ \varepsilon _{g}+E_{1}(k)\alpha _{k}^{\dagger
}\alpha _{k}+E_{2}(k)\beta _{k}^{\dagger }\beta _{k}\right]  \label{diag}
\end{equation}
where the coefficients $u_{k}$, $v_{k}$ are constrained by the condition $%
u_{k}^{2}-v_{k}^{2}=1$, $\alpha _{k}$ and $\beta _{k}$ are the normal mode
boson operators, $E_{1}(k)$ and $E_{2}(k)$ are the energies of the two
excitation modes and $\varepsilon _{g}$ is the ground state energy per site,
with $u(k)$ and $v(k)$ defined as 
\begin{mathletters}
\label{dia}
\begin{eqnarray}
u(k) &=&\sqrt{\frac{A_{1}+A_{2}+\xi _{k}}{2\xi _{k}}} \\
v(k) &=&\sqrt{\frac{A_{1}+A_{2}-\xi _{k}}{2\xi _{k}}} \\
\xi _{k} &=&\sqrt{\left( A_{1}+A_{2}\right) ^{2}-4B(k)^{2}}
\end{eqnarray}
The two excitation modes are 
\end{mathletters}
\begin{mathletters}
\label{dia}
\begin{eqnarray}
E_{1}(k) &=&\left( A_{1}-A_{2}+\xi _{k}\right) /2  \label{mode1} \\
E_{2}(k) &=&\left( A_{2}-A_{1}+\xi _{k}\right) /2  \label{mode2}
\end{eqnarray}
It is easy to see that $E_{1}$ is the gapless mode and $E_{2}$ has a gap.%
\newline

The ground state energy per site $\varepsilon _g$ is given by 
\end{mathletters}
\begin{equation}
\varepsilon _g=C-A_1-A_2+\sum_k\xi _k  \label{enr}
\end{equation}
and the staggered magnetization in the two sub-lattices corresponding to the
spins $s_1$ and $s_2$ respectively is 
\begin{mathletters}
\label{mag}
\begin{eqnarray}
\ &&M_1=S_1-<D> \\
\ &&M_2=<D>-S_2
\end{eqnarray}
with $<D>=<a_i^{\dagger }a_i>=<b_j^{\dagger }b_j>$ is the average taken in
the ground state, which is the Neel state, at zero temperature. This average
of spin deviation operators can be determine from 
\end{mathletters}
\begin{equation}
<D>=\frac 1N\sum_kv^2(k)
\end{equation}
with $k$ running over half the Brillouin zone.\\

For the three kinds of alternating spin chains, referred to as $(1$,$\frac 12%
),$ $(\frac 32,\frac 12)$ and $(\frac 32,1)$, the ground state energy, the
excitation energies and magnetization can now be calculated as functions of
the dimerization parameter $\delta $ . Previous calculations invariably took
spin-spin exchange couplings alternately as $J(1\pm \delta )$, which, as
mentioned above, can be taken as an expansion of the interaction in Eq.(\ref
{fullj}) to the first order in $\delta $, implying that the results are
valid only in the critical regime $\delta \rightarrow 0$. The advantage of
taking the variable nearest neighbor exchange coupling is that the results
will then be valid also in the limit $\delta \rightarrow 1$.\newline

The ground state energies $\varepsilon _g$ per site for undimerized chains, $%
\delta =0,$ were found, expectedly, to be the same as those found earlier%
\cite{pati} for the three systems. After including dimerization, our
calculations also confirm that the ground state energy of all the three
systems described above decreases with $\delta $. This is shown in Figure 2,
where energy gain $\varepsilon _g(\delta )-\varepsilon _g(0)$ is plotted
against $\delta $. Numerical fitting shows that, as against earlier results,
the magnetic energy gain $\varepsilon _g(\delta )-\varepsilon _g(0)$ has a
logarithmic dependence on $\delta $, $\frac{\delta ^\nu }{\left| \ln \delta
\right| }$, for the three systems discussed here, with values of $\nu $
between $1.4-1.6$ in the entire range $0\leq \delta <1$. Fig. 2 shows that
the chain $(\frac 32,1)$ has higher gain than the other two systems. 
\newline

As expected, our calculations also find two branches of the excitation
spectrum, one gapless and the other with a gap at $k=0$, in the three
systems.\\

As stated above, the LSW theory with the spin-spin exchange coupling $J(1\pm
\delta )$ in the presence of dimerization allowed no $\delta -$dependence of
the energy gap in the second mode, while the DMRG found almost a linear $%
\delta -$dependence for these spin systems\cite{swapan,pati}. We find that
by including a variable nearest neighbor coupling constant defined in Eq.(%
\ref{fullj}) the spin wave theory also allows for a $\delta -$dependent
energy gap $\Delta (\delta )=E_2(\delta )-\varepsilon _g(\delta )$ in the
second mode. This is because the $\delta $-dependent terms do not now cancel
out for the gapped excitation mode at $k=0$ as they did with the coupling $%
J(1\pm \delta )$. The dependence is found to follow the same logarithmic
behavior, $\frac{\delta ^\nu }{\left| \ln \delta \right| },$ as the ground
state energy, with $\nu $\ varying between $1.4-1.6$ for the entire range $%
0\leq \delta <1$. This is true for all the three spin systems defined here,
and is larger for the $(\frac 32,1)$ system than the other two (see figure
3).\newline
.

The staggered magnetization $M(\delta )$ was also found to follow the
logarithmic $\delta $-dependence up to $\delta \leq 0.5$, but follows a
different behavior for $\delta >0.5$. $M_1(\delta )$ against dimerization is
shown in Fig.(4) for the three systems. Again we see from Fig.(4) that the
chain with $(\frac 32,1)$ has higher value of magnetization than the other
two systems.\newline

It is worth mentioning here that by using coupled cluster method\cite{aiman}%
, we had found that the ground state energy and the staggered magnetization
of a spin-half Heisenberg chain follow the same logarithmic behavior using
the variable exchange coupling defined in Eq.(\ref{fullj}), in both small
and large values of dimerization $\delta $. This gives us more confidence
about the results we have obtained by LSW theory.\newline

\section{Two Dimensional Alternating Hamiltonian :-}

The dimerization on two dimensions lattices differs from that on chains.
There are several ways in which distortions of a square lattice can occur,
each one of the possible configurations giving a different dependence of the
ground state energy on the dimerization parameter\cite{aiman}.\\

We will use some of these configurations, illustrated in Fig. (5), to study
the alternating spin square lattices. The alternating dimerized Hamiltonian
for a two dimensional system can be written in general as

\begin{equation}
H=\sum_{i,j}^{\sqrt{N}}\sum_{\mu =\pm 1}\left[ J_{x,\mu }{\bf S}%
_{1,i,j}\cdot {\bf S}_{2,i+\mu ,j}+J_{y,\mu }{\bf S}_{1,i,j}\cdot {\bf S}%
_{2,i,j+\mu }\right]  \label{2dalt}
\end{equation}
where the indices $1$ and $2$ on the spin vectors refer to the two
sublattices with spins of magnitude $s_1$ and $s_2$. For reasons described
earlier, we use variable nearest neighbor exchange couplings. These are
defined for different configurations as follows:

{\bf Configuration (a)}\newline

$J_{x,\mu }=\frac J{(1+\mu \delta )}\simeq J(1-\mu \delta )$

$J_{y,\mu }=J.$\newline

{\bf Configuration (b)}\newline

$J_{x,\mu }=\frac J{(1+\mu \delta )}\simeq J(1-\mu \delta )$

$J_{y,\mu }=\frac J{\sqrt{1+\delta ^2}}\simeq J(1-\frac{\delta ^2}2)$%
\newline

{\bf Configuration (c)}\newline

$J_{x,\mu }=J_{y,\mu }=\frac J{(1+\mu \delta )}\simeq J(1-\mu \delta )$ 
\newline

{\bf Configuration (d)}\newline

$J_{x,\mu }=\frac J{(1+\mu \delta )}\simeq J(1-\mu \delta )$\newline

$J_{y,\mu }=\frac J{\sqrt{\delta ^2+\left( 1+\mu \delta \right) ^2}}\simeq
J\left( 1-\mu \delta -(1-\frac{\mu ^2}2)\delta ^2\right) $\newline

{\bf Configuration (e)}\newline

$J_{x,\mu }=J_{y,\mu }=\frac J{\sqrt{\delta ^2+\left( 1+\mu \delta \right) ^2%
}}\simeq J\left( 1-\mu \delta -(1-\frac{\mu ^2}2)\delta ^2\right) $\newline

We would like to investigate the five configurations resulting from
dimerization of a square lattice in order to see (i) which one of these
leads to the largest gain in magnetic energy as the dimerization sets in,
(ii) if the use of variable exchange coupling leads to a single power law
behavior valid for the entire range of $\delta $, and how the law differs
from that in the case of chains$,$ (iii) the $\delta $-dependence of the
second mode of excitation $E_2$, (iv) the behavior of staggered
magnetization, and (v) the generality of these investigations regarding the
three spin systems discussed here.\newline

The linear spin wave analysis follows the same procedure as for the chain
above. The same equations are applicable in this case, but the various
coefficients entering the theory have now the following values:\vspace{1pt} 
\begin{mathletters}
\label{mag}
\begin{eqnarray}
A_1 &=&J_ps_2 \\
A_2 &=&J_ps_1 \\
B(k) &=&\Gamma (k)\sqrt{s_1s_2} \\
C &=&-J_ps_1s_2
\end{eqnarray}
where 
\end{mathletters}
\[
\Gamma (k)=\sqrt{\left( J_{px}\cos (k_x)+J_{py}\cos (k_y)\right) ^2+\left(
J_{mx}\sin (k_x)+J_{my}\sin (k_y)\right) ^2} 
\]
and

$J_{p}=(J_{x,+1}+J_{x,-1}+J_{y,+1}+J_{y,-1})/4,$

$J_{px}=(J_{x,+1}+J_{x,-1})/4,$

$J_{py}=(J_{y,+1}+J_{y,-1})/4,$

$J_{mx}=(J_{x,+1}-J_{x,-1})/4,$

$J_{my}=(J_{y,+1}-J_{y,-1})/4$.

\vspace{1pt}

The ground state energy $\varepsilon _g(\delta )$ defined in Eq.(\ref{enr}),
energies of the two excitation modes $E_i(k)$ in Eq.(\ref{dia}) and
staggered magnetization $M(\delta )$ defined in Eqs.(\ref{mag}) can now be
calculated as functions of the dimerization parameter $\delta .$\\

The ground state energy $\varepsilon _g(\delta =0)$ is found to be $%
-1.2,-1.7158$ and $-3.3709$ for the three spin systems $(1,\frac 12)$, $(%
\frac 32,\frac 12)$ and $(\frac 32,1)$ respectively. Staggered magnetization 
$M_1\{M_2\}$ on the first \{second\} sublattice is $0.8907$ $\{-0.3907\}$, $%
1.4241$ $\{-0.4241\}$ and $1.3597$ $\{-0.8597\}$ for the three systems .
These values are listed in Table 2. \newline

Our calculations confirm that, like in chain, the gain in magnetic energy
increases with $\delta $ in all the proposed configurations. This is shown
in Figure 6, where the energy gain $\varepsilon _g(\delta )-\varepsilon
_g(0) $ is plotted against $\delta $ for the five configurations. It also
shows that the plaquette configuration of Fig. 5(c) is energetically the
most favorable state, while there is hardly a discernible difference among
the configurations (a), (b) and (d). It is also interesting to note that the
magnetic energy gain under dimerization of an alternating spin square
lattice also varies as $\frac{\delta ^v}{\left| \ln \delta \right| }$ with $%
\nu =1.4-1.6$ in the entire range 0$\leq \delta <1$, exactly as in the case
of a chain. This is singularly an effect of taking \ the variable exchange
coupling defined above.\\

The $\delta $ dependence of the energy gap, $\Delta (\delta )=E_2(\delta
)-\varepsilon _g(\delta )$, for the five configurations is shown in Fig. 7,
showing greater stabilization of the dimerized state with increasing $\delta 
$. We also find that, like the magnetic energy gain, the energy gap
increases with $\delta $ as $\ \frac{\delta ^v}{\left| \ln \delta \right| }$
in the small $\delta $ regime for all the five configurations with $\nu
=1.4-1.6$. The configurations (a) - (d) also have the same dependence on $%
\delta $ in the entire range of $\delta .$ The difference between the
dimerization of a square lattice for these configurations is again markedly
brought out in Fig. 7.\\

Our calculations give staggered magnetization for the un-dimerized
alternating spin square lattice $M_1(\delta =0)=0.8907$, $1.4241$ and $%
1.3597 $ and $M_2(\delta =0)=-0.3907$, -$0.4241$ and $-0.8597$ for $(1,\frac 
12)$, $(\frac 32,\frac 12)$ and $(\frac 32,1)$ respectively. As dimerization
sets in, magnetization decreases in all the configurations we have chosen,
as shown in Fig. 8. This is also the case for the entire range $0\leq \delta
<1$, except in the case of configuration (e) for which the magnetization
rises again after $\delta >\frac 12$.\newline

Configuration (e) is peculiar in the sense that $\delta =\frac 12$ is a
special point for it; the shorter bond length is symmetric about this point,
having a minimum value of $\frac 1{\sqrt{2}}$. At this point the distortions
give rise to a rectangular lattice with sides $\sqrt{2}$ and $\frac 1{\sqrt{2%
}}$. The energy gain increases with $\delta $ up to $\delta =\frac 12$, and
then decreases.\\

For all the five configurations, we found that the magnetization also varies
as $\frac{\delta ^\nu }{\left| \ln \delta \right| }$ in the small $\delta $
regime with the exponent $\nu =1.4-1.6$, exactly as the energy gain and the
energy gap. However, while for configurations (a-d) in the full range $0\leq
\delta <1$ the magnetization follows the same power law with the exponents
as $\nu =1.4-1.6$, configuration (e) has a distinctly different behavior in
this regime. \newline

In summary, we have studied the spin-Peierls dimerization of an alternating
spin Heisenberg system on a chain and a square lattice taking variable
exchange couplings based on the ansatz $J(a)=\frac Ja,$ for three kind of
alternating spin systems, namely $(1,\frac 12)$, $(\frac 32,\frac 12)$ and $(%
\frac 32,1)$. We have included different possibilities of dimerization in
the case of a square lattice. The ground state energy as well as staggered
magnetization decrease continuously with increasing dimerization in both 1D
and 2D . In 2D, the plaquette configuration with dimerization taking place
simultaneously along both the principal square axes has markedly lower
ground state energy and magnetization than the other configurations; (a),
(b), (d) and (e). The plaquette configuration stands out as the most favored
mode of dimerization. The energy gap also corroborates the above
conclusions. It has also been shown that the magnetic energy gain, energy
gap and staggered magnetization follow a uniform dependence upon the
dimerization parameter $\delta $ as $\frac{\delta ^v}{\left| \ln \delta
\right| }$ (i) in chains as well as in square lattices; (ii) in systems
consisting of different pairs of spins $s_1$ and $s_2$; (iii) for the
magnetic energy gain, the energy gap, the energy of the gapped magnetic
excitation mode as well as for the sublattice magnetization; (iv) for all
the configurations of the square lattice; and (v) in the entire range $0\leq
\delta <1$. The variable exchange coupling also allows \ the energy of the
gapped excitation spectrum to be $\delta $-dependent even in the linear spin
wave theory.\\

We would like to thank to Dr. D. Sen for providing Refs. 5 and 6.

\newpage\ 

{\bf Figure captions}\newline

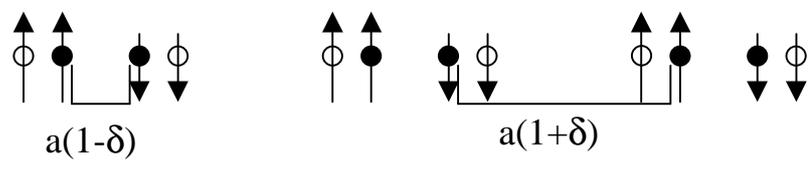
Figure 1: A schematic sketch of an alternating spin chain. The larger and
smaller arrows indicate the larger ($s_1)$ and the smaller ($s_2)$ spins.
The hollow (filled) circles represent the positions of spins in the
undisturbed (dimerized) chain.\newline

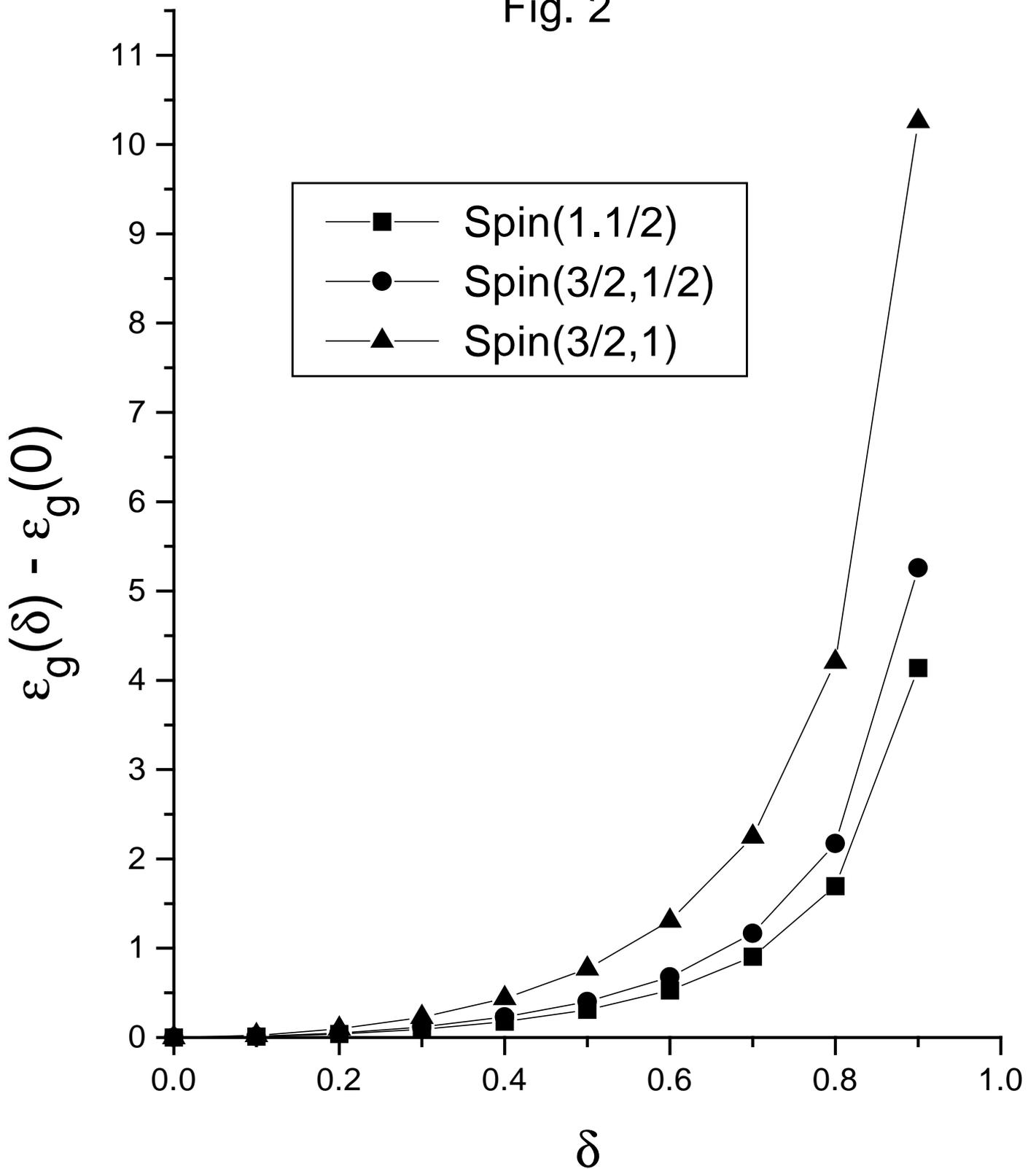
Figure 2: The magnetic energy gain $\varepsilon _g(\delta )-\varepsilon
_g(0) $ vs the dimerization parameter $\delta $ for 1D alternating spin
chain in the full range of the dimerization parameter $0\leq \delta <1$.%
\newline

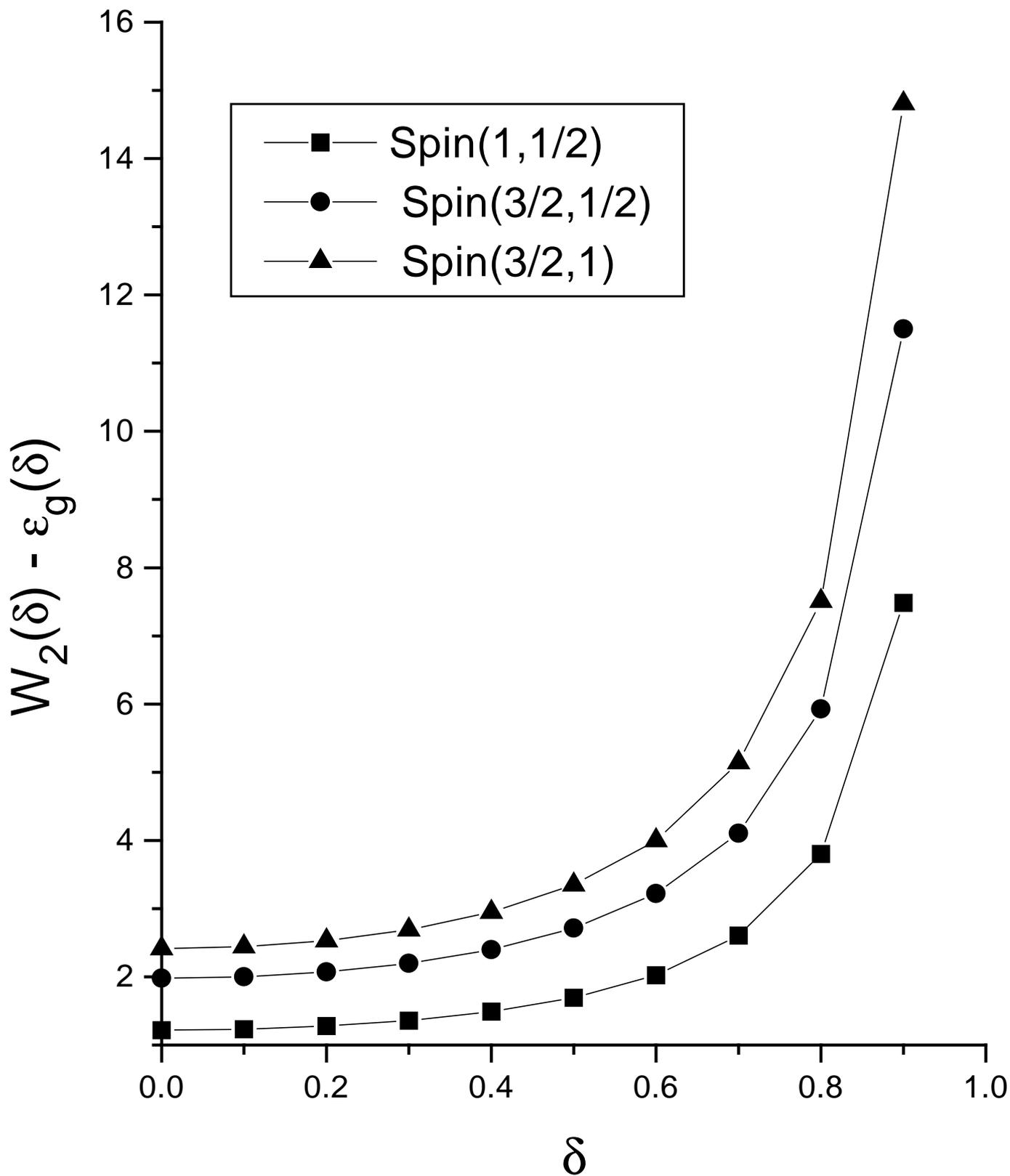
Figure 3: $\delta $-dependence of the energy gap as dimerization sets in for
different alternating spin chains.\newline

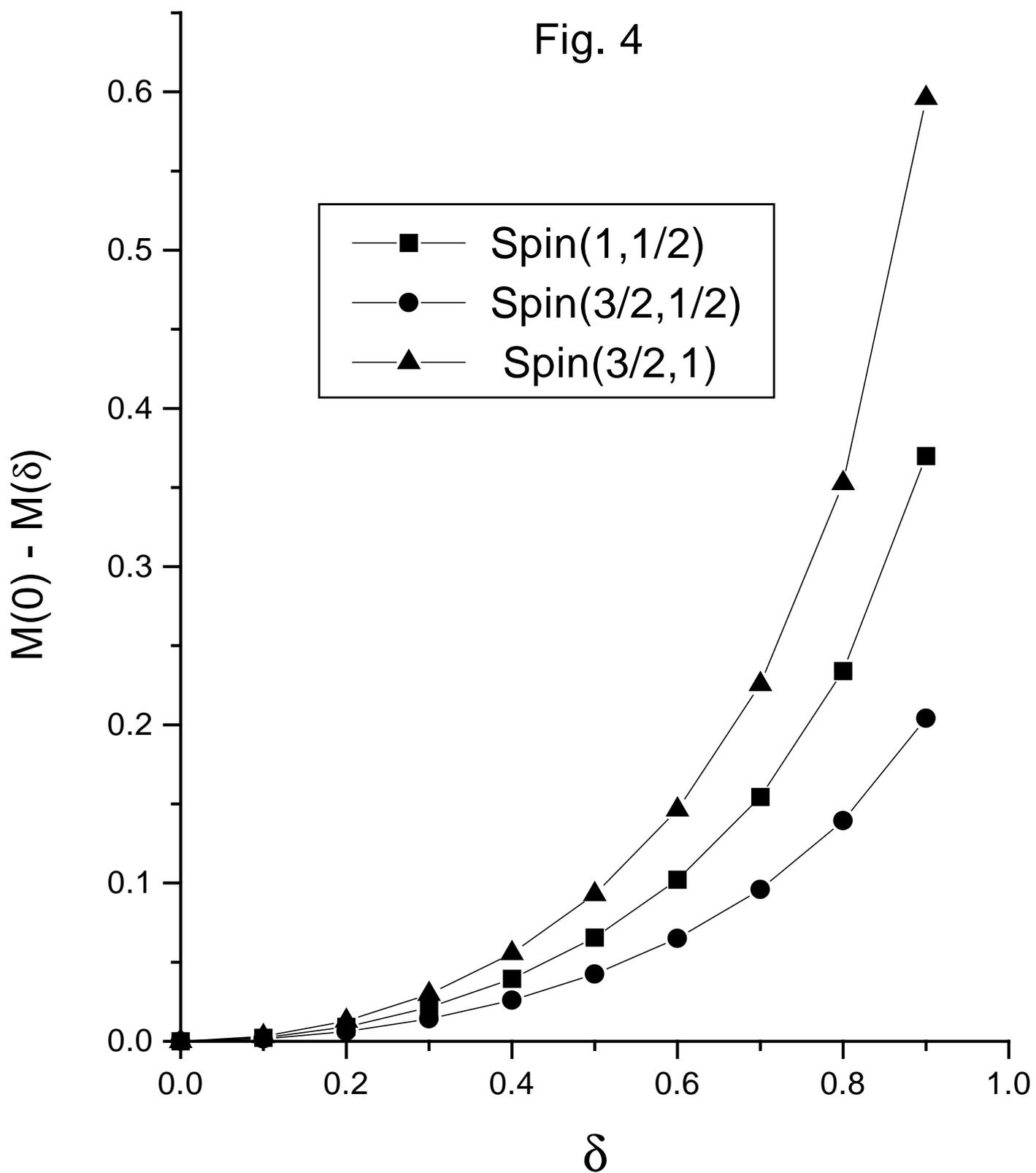
Figure 4: $\delta $-dependence of the staggered magnetization for one of the
two sublattices, $S_1$, for the three alternating spin chains.\newline

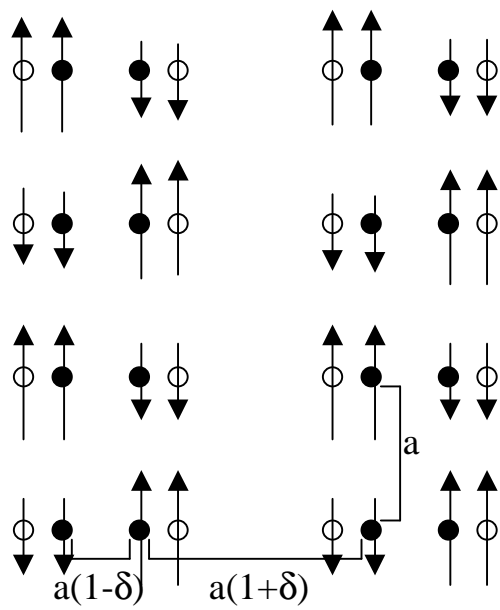
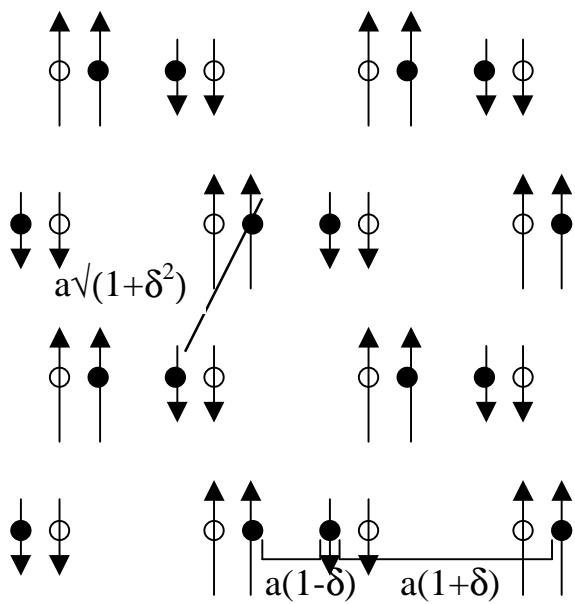
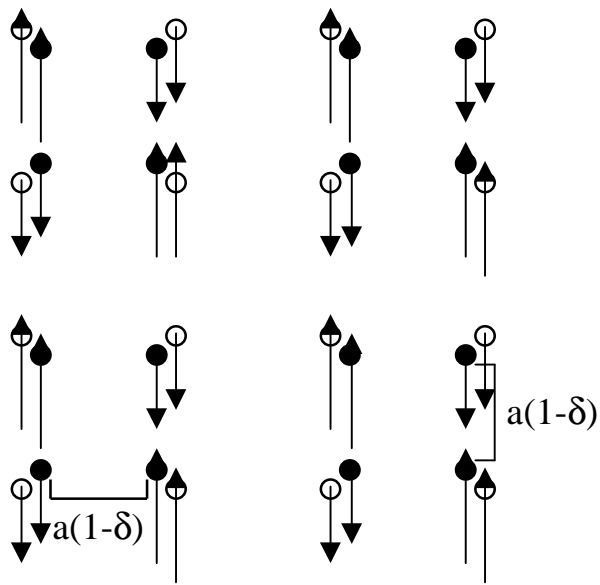
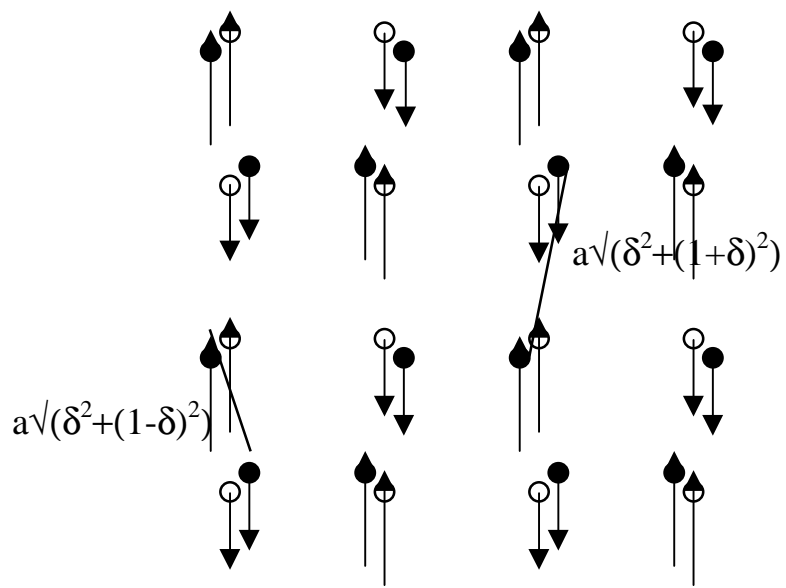
Figure 5: Five configurations for the dimerization of a square lattice. (a)
a columnar configuration caused by a longitudinal $(\pi ,0)$ static phonon;
(b) a staggered configuration caused by a $(\pi ,\pi )$ static phonon with
polarization along{\bf \ }$x$-direction. Like (a), the dimerization occurs
along one direction only, but the sequence of alternate couplings itself
alternates along the other direction; (c) dimerization along both the
directions, caused by $(\pi ,0)$ and $(0,\pi )$ phonons, making a plaquette
of four nearest neighbour spins; (d) again, dimerization along both the
directions, but taken staggered along the vertical direction; {\bf \ }(e)
another staggered dimerization that is caused by a longitudinal $(\pi ,\pi )$
phonon. Large arrow belongs to the first sublattice and short one belongs to
second sublattice. And the open circles indicate the square lattice sites
and the solid ones show the dimerized lattice.\newline

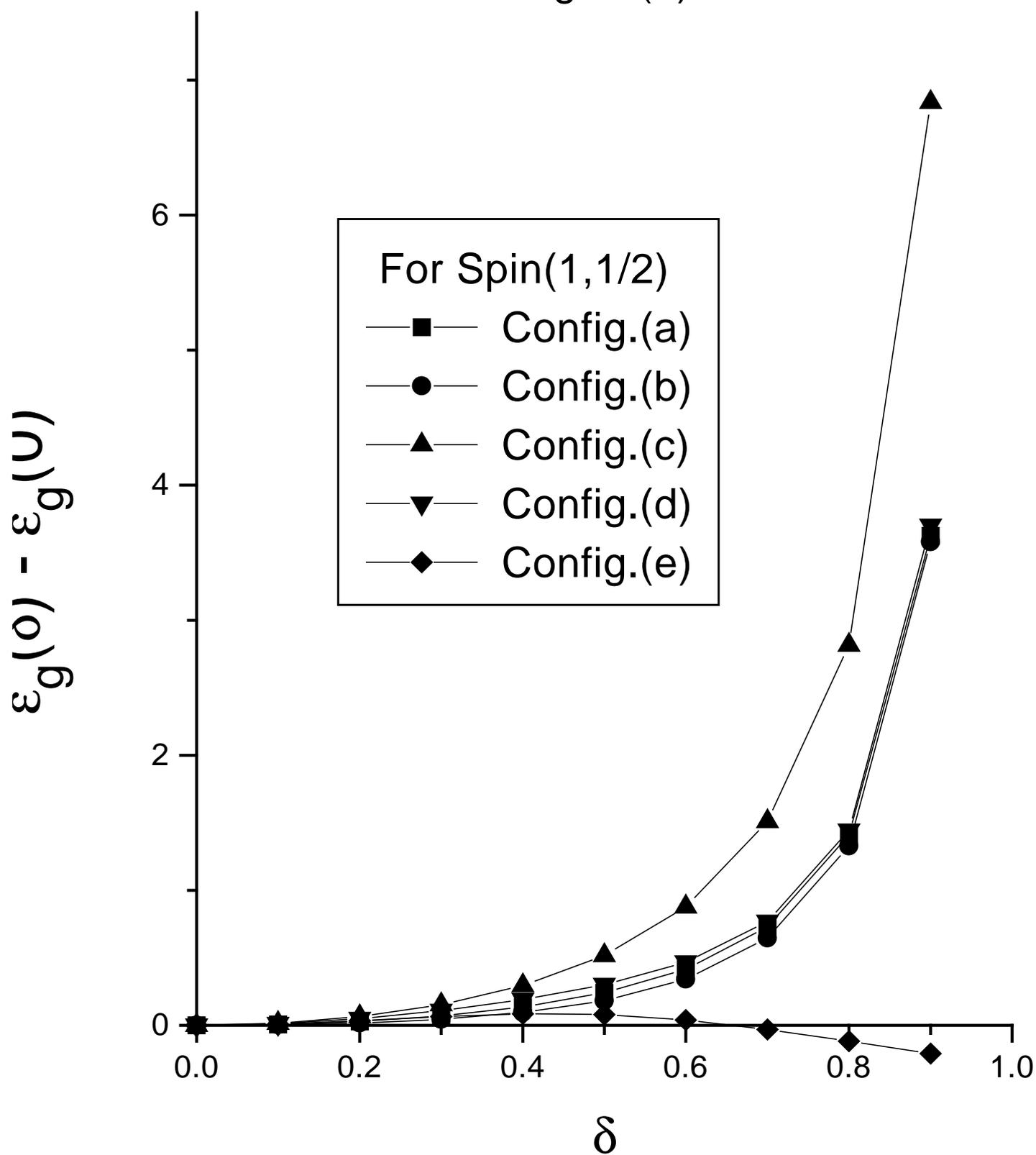
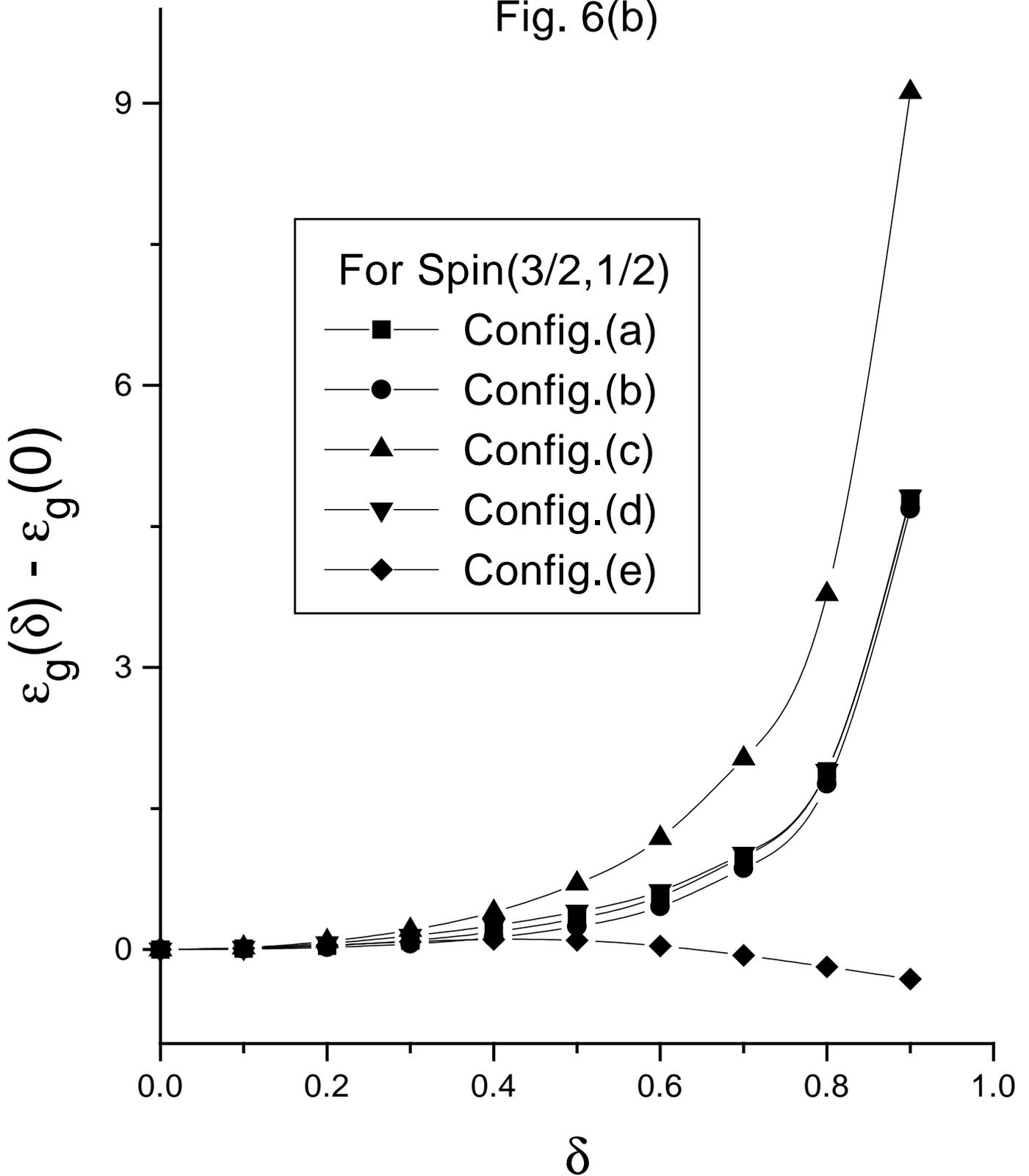
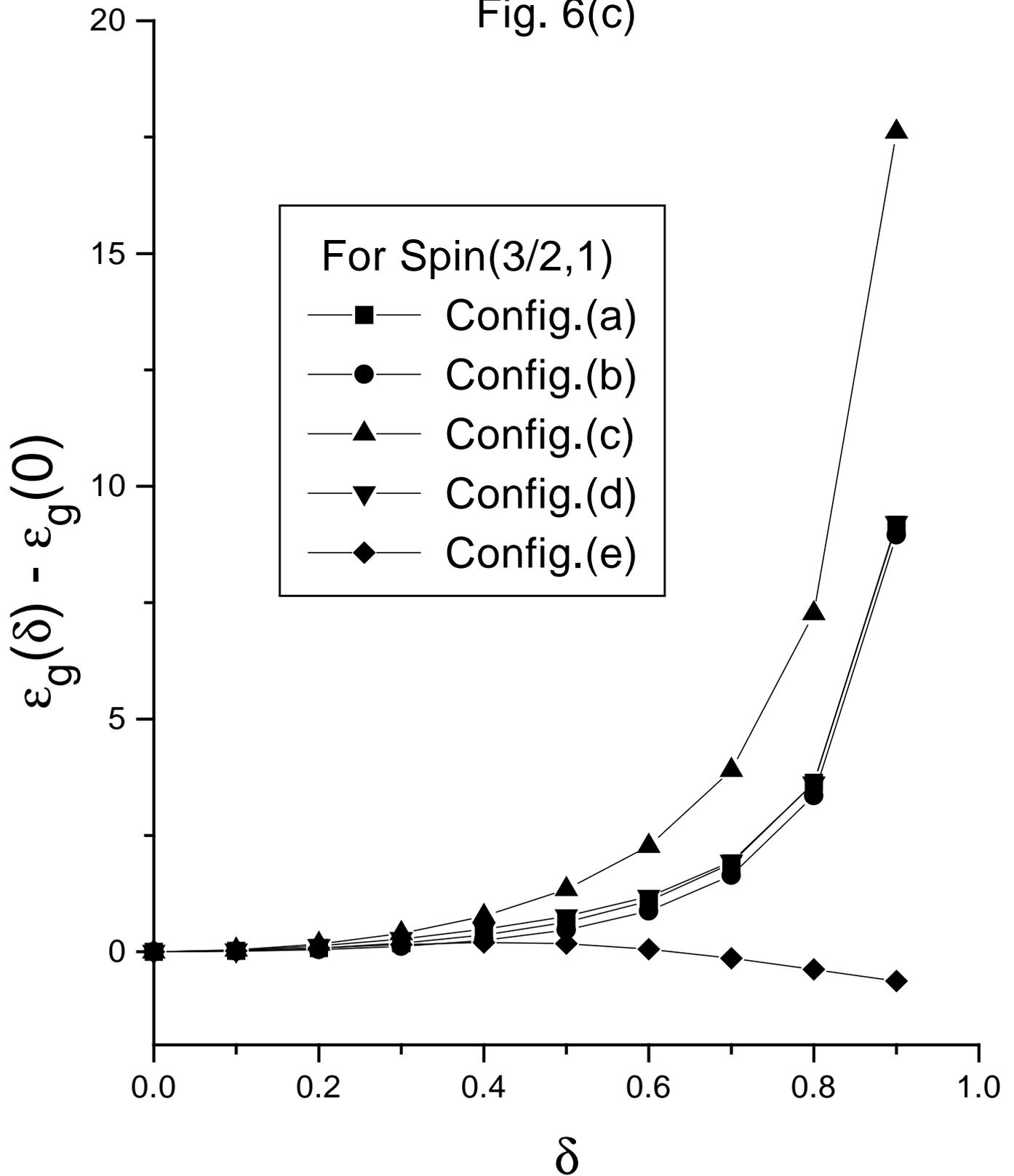
Figure 6: The gain in magnetic energy $\varepsilon _g(\delta )-\varepsilon
_g(0)$ as dimerization sets in with increasing $\delta $ for the five
configurations of a square lattice in the range $0\leq \delta <1$ for (a)
spin $(1,\frac 12)$, (b) spin $(\frac 32,\frac 12)$ and (c) spin $(\frac 32%
,1)$.\newline

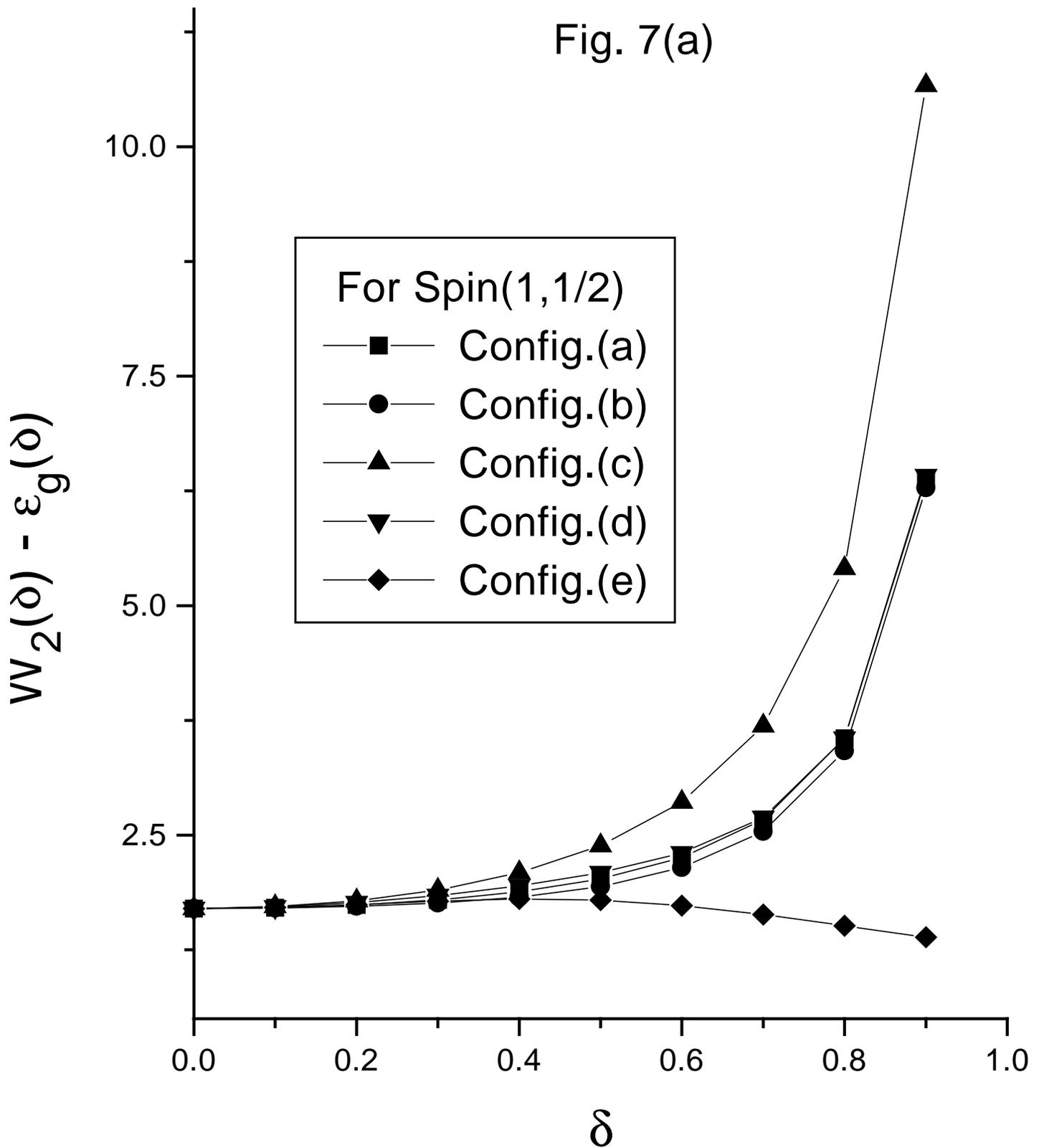
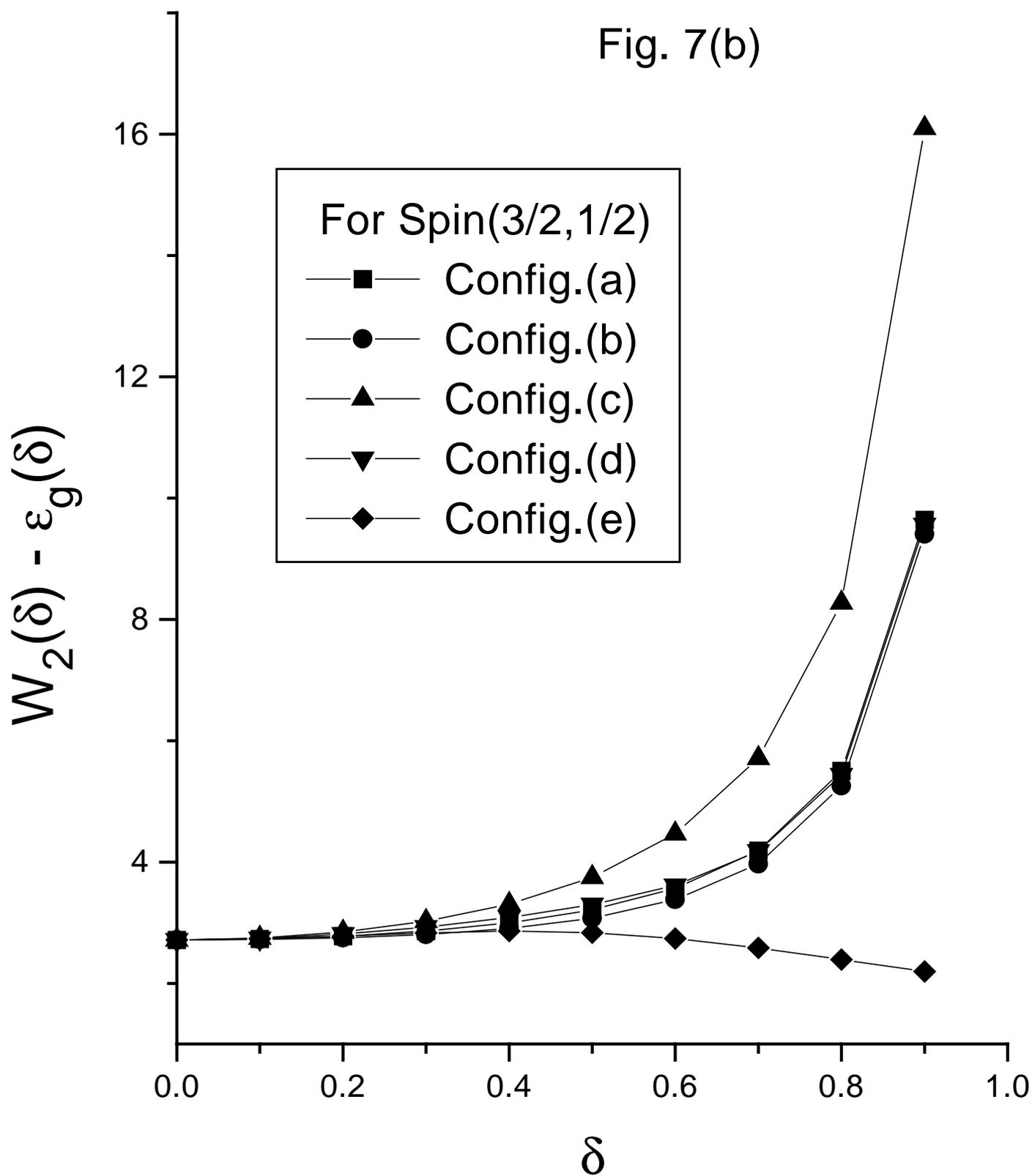
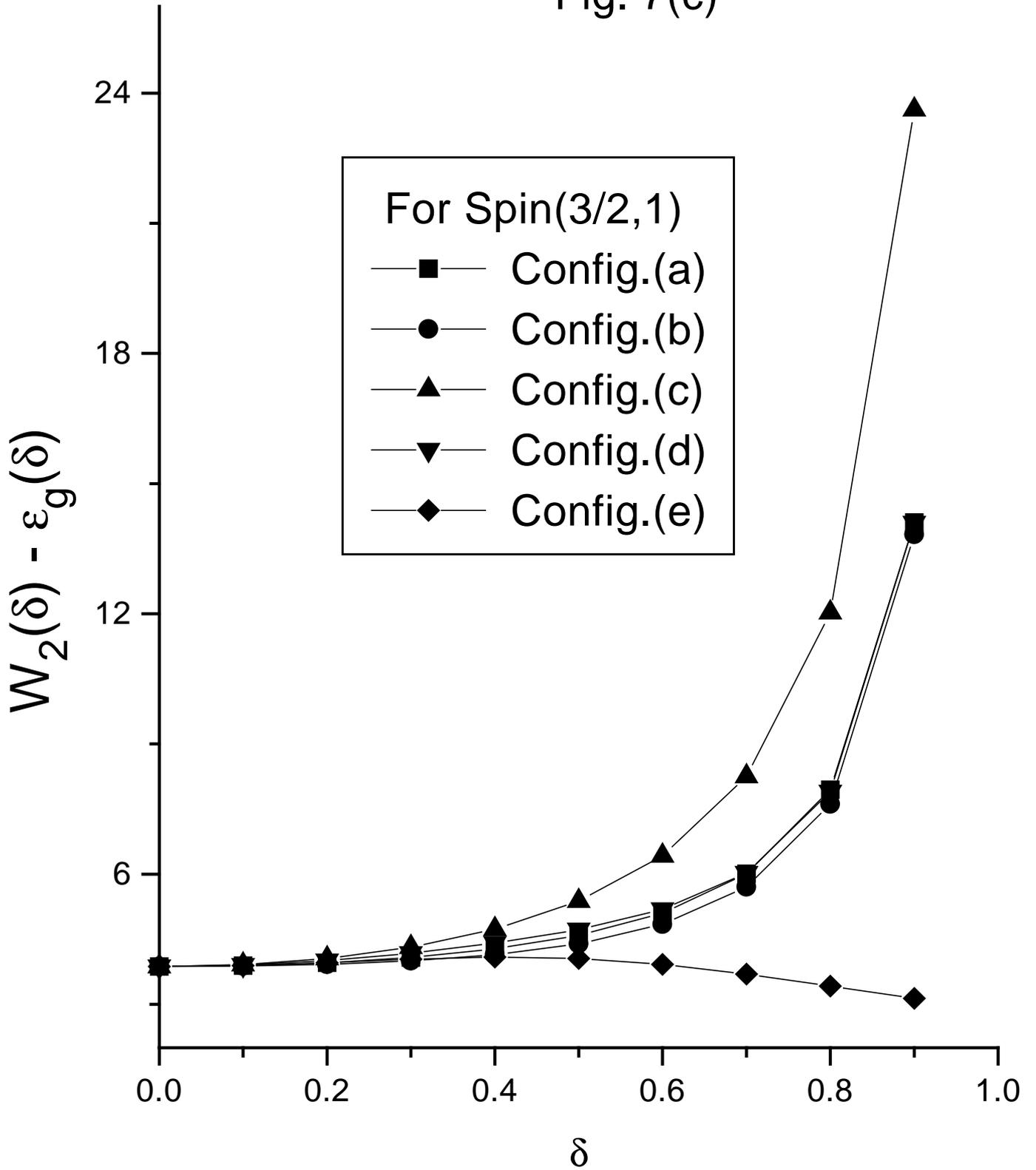
Figure 7: Dependence of the energy gap $\Delta $ on $\delta $ for the five
dimerization configurations of the alternating square lattices for (a) spin $%
(1,\frac 12)$, (b) spin $(\frac 32,\frac 12)$ and (c) spin $(\frac 32,1)$.%
\newline

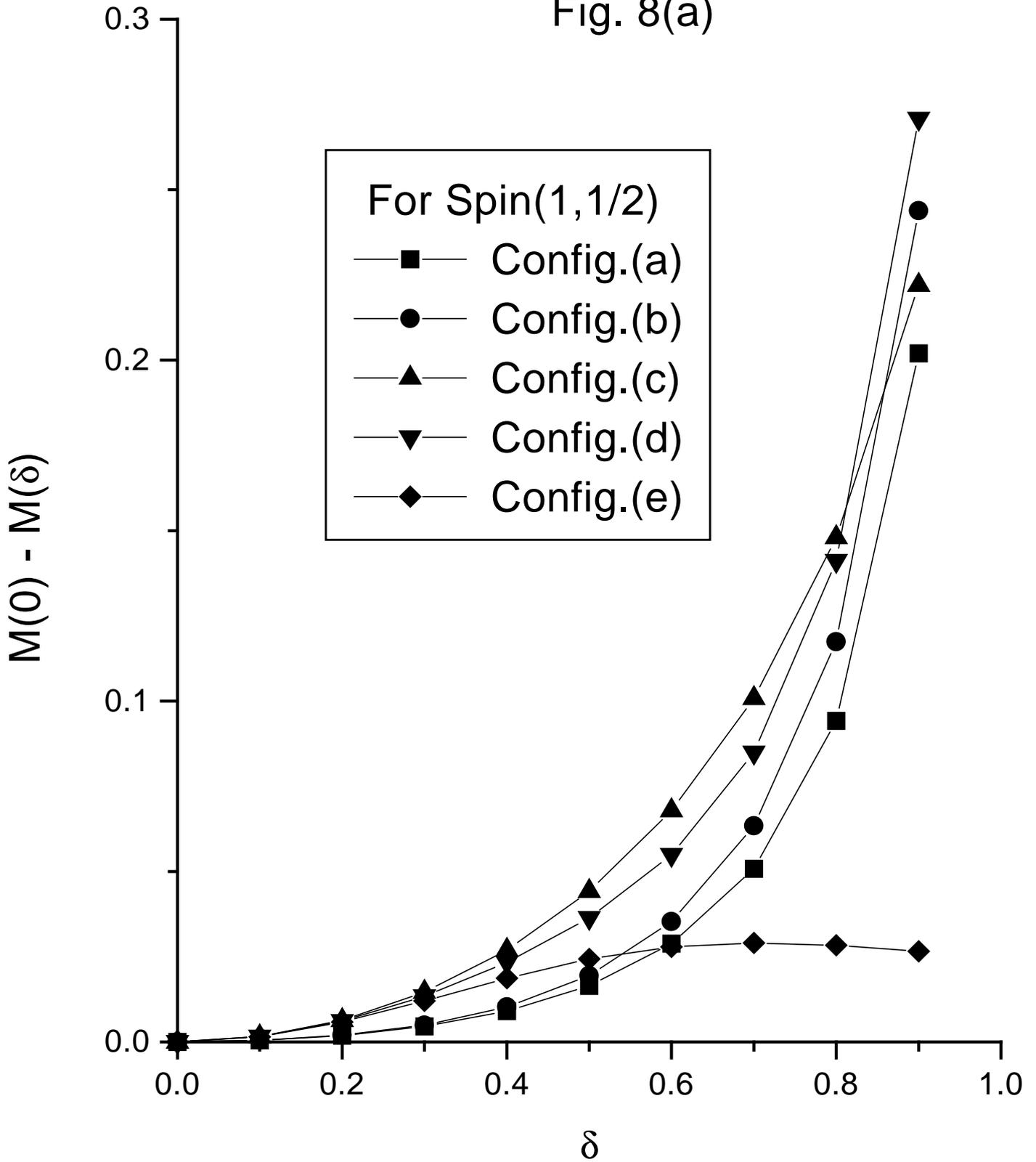
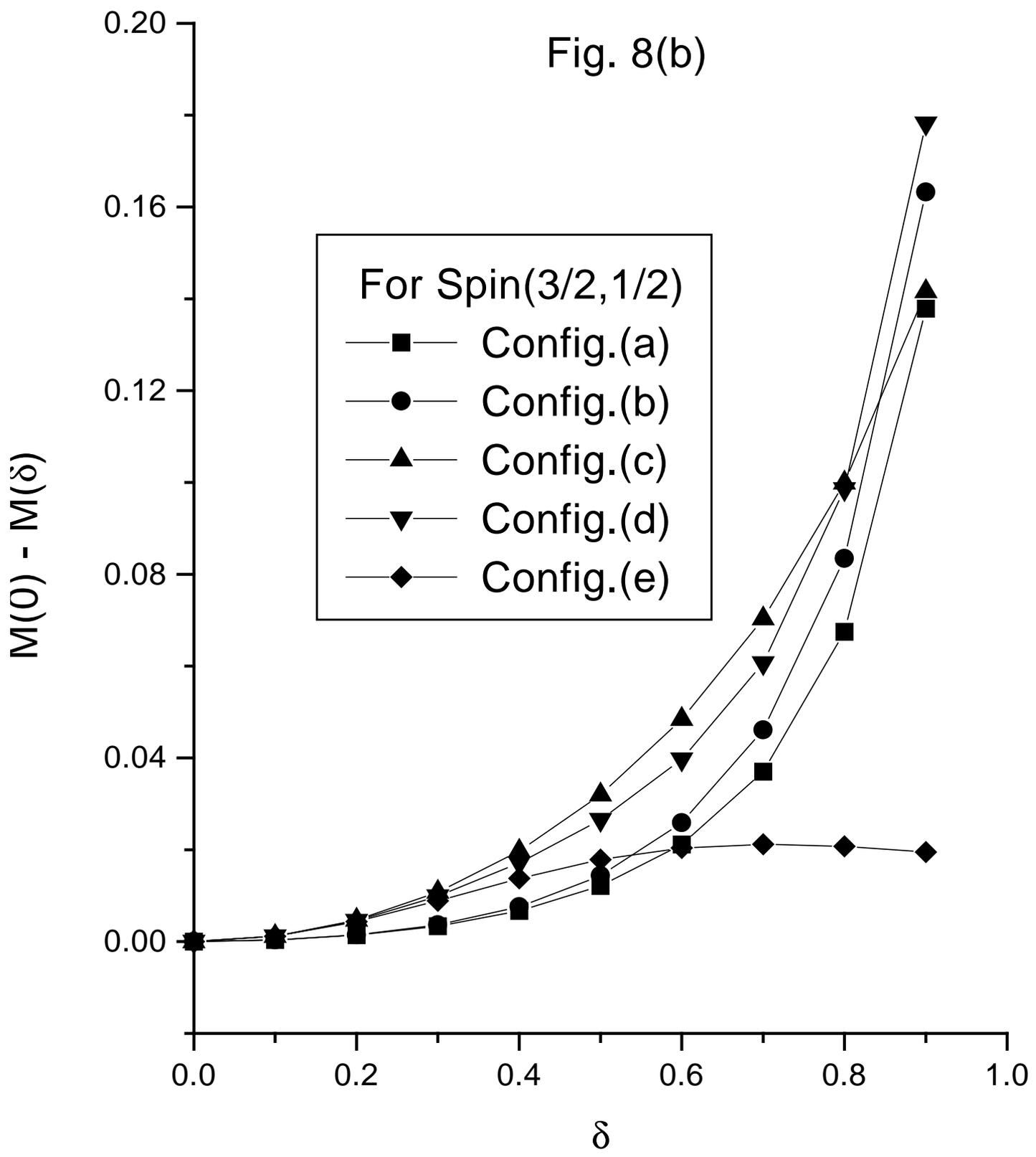
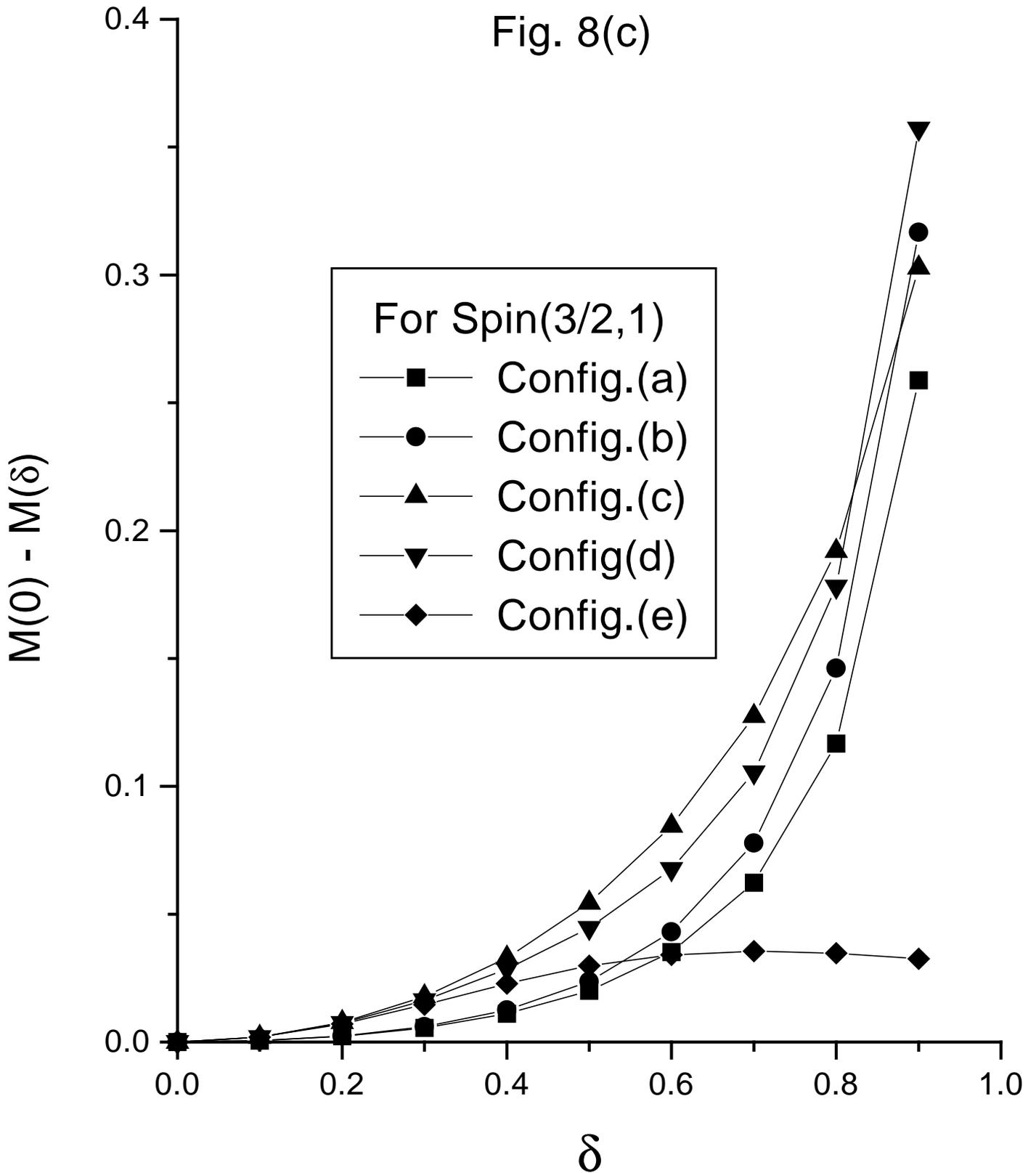
Figure 8: $\delta $-dependence of the staggered magnetization of an
alternating spin square lattice calculated for the five dimerization
configurations. (a) spin $(1,\frac 12)$, (b) spin $(\frac 32,\frac 12)$ and
(c) spin $(\frac 32,1)$.\newline

\newpage

Table 1: Summary of the ground state energy per site and sublattice
magnetization values calculated by different methods for alternating spin
chains made up of the three spin systems namely $(1,\frac 12)$, $(\frac 32,%
\frac 12)$ and $(\frac 32,1).$\\\flushleft 

\begin{tabular}{||c|c|c|c|c||}
\hline\hline
Spin & Method & $\varepsilon _g$ & $M_1$ & $M_2$ \\ \hline\hline
$(1,\frac 12)$ & MP$^{\text{\cite{kolezhuk}}}$ & $-07245$ & $0.779$ & $%
-0.279 $ \\ \cline{2-5}
& QMC$^{\text{\cite{kolezhuk,brehmer}}}$ & $-0.7275$ & $0.793$ & $-0.293$ \\ 
\cline{2-5}
& LSWT$^{\text{\cite{brehmer,pati}}}$ & $-0.718$ & $0.695$ & $-0.195$ \\ 
\cline{2-5}
& DMRG$^{\text{\cite{pati}}}$ & $-0.72709$ & $0.79428$ & $-0.29248$ \\ 
\cline{2-5}
& SWE$^{\text{\cite{ivanov,ivanov2}}}$ & $-0.72715$ & $0.79388$ &  \\ 
\cline{2-5}
& MSW$^{\text{\cite{yamamoto3}}}$ & $-0.7295$ &  &  \\ \hline
$(\frac 32,\frac 12)$ & LSWT$^{\text{\cite{pati}}}$ & $-0.979$ & $1.315$ & $%
-0.314$ \\ \cline{2-5}
& DMRG$^{\text{\cite{pati}}}$ & $-0.98362$ & $1.35742$ & $-0.35742$ \\ 
\cline{2-5}
& SWE$^{\text{\cite{ivanov,ivanov2}}}$ & $-0.9834$ & $1.3666$ &  \\ \hline
$(\frac 32,1)$ & LSWT$^{\text{\cite{pati}}}$ & $-1.914$ & $1.040$ & $-0.540$
\\ \cline{2-5}
& DMRG$^{\text{\cite{pati}}}$ & $-1.93096$ & $1.14427$ & $-0.644$ \\ 
\cline{2-5}
& SWE$^{\text{\cite{ivanov,ivanov2}}}$ & $-1.9316$ & $1.1461$ &  \\ 
\hline\hline
\end{tabular}
\vspace{1.0in}

Table 2: The ground state energy per site and the staggered magnetization of
the un-dimerized alternating spin square lattice for the three spin systems
as calculated in the linear spin wave theory.\newline
\flushleft 

\begin{tabular}{||c|c|c|c||}
\hline\hline
Spin system & $\varepsilon _g$ & $M_1$ & $M_2$ \\ \hline\hline
$(1,\frac 12)$ & $-1.2$ & $0.8907$ & $-0.3907$ \\ \hline
$(\frac 32,\frac 12)$ & $-1.7158$ & $1.4241$ & $-0.4241$ \\ \hline
$(\frac 32,1)$ & $-3.3709$ & $1.3597$ & $-0.83597$ \\ \hline\hline
\end{tabular}
\vspace{1.0in}

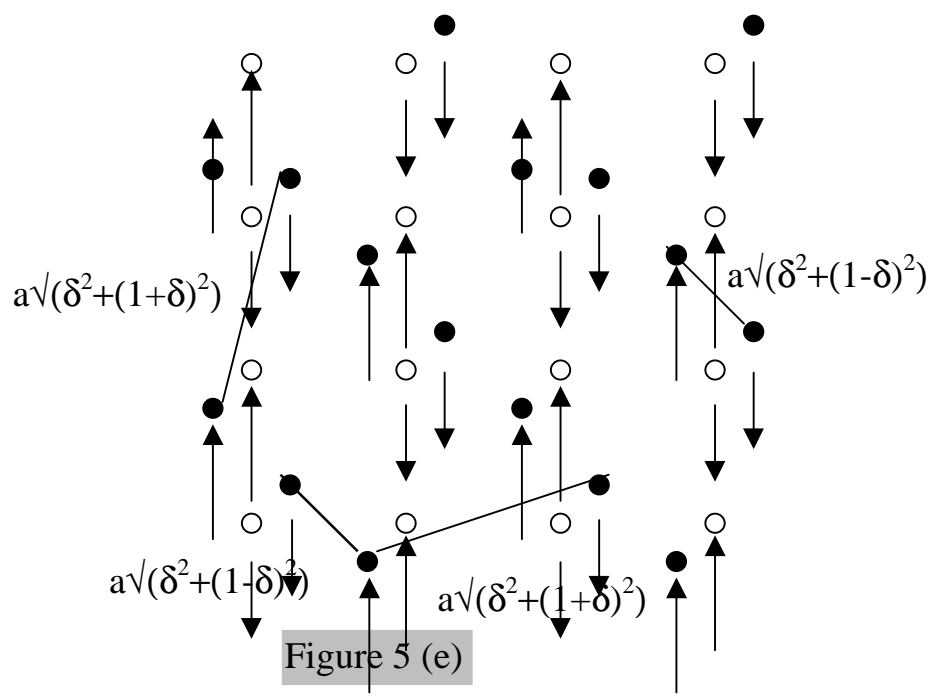
\end{document}